\documentstyle [twoside,12pt]{article}
\begin{document}
\newcommand{\de}{\mbox{$ {\rm d}$}}
\newcommand{\T}{\mbox{${\rm T}$}}
\sloppy
\setlength{\baselineskip}{.7cm}
\sloppy
\begin{center}
{\bf Integrable nonlinear field equations and loop algebra
structures}
\end{center}
\begin{center}
{\bf E. Alfinito, M. Leo, R.A. Leo, M. Palese and G. Soliani }
\end{center}
\begin{center}
{\it Dipartimento di Fisica dell'Universit\`a, 73100 Lecce, Italy\\
 and Istituto Nazionale di Fisica Nucleare, Sezione di Lecce. }
\end{center}
\medskip
\medskip
\medskip
\medskip

\vspace{0.5cm}

\begin{abstract}
We apply the (direct and inverse) prolongation method to a couple of
nonlinear Schr{\"o}dinger equations. These are taken as a laboratory
field model for analyzing the existence of a connection between
the integrability property and loop algebras. Exploiting a
realization of the Kac-Moody type of the incomplete prolongation
algebra associated with the system under consideration, we develop a
procedure with
allows us to generate a new class of integrable nonlinear field
equations containing the original ones as a special case.
\end{abstract}


\section{Introduction}
The existence of Kac-Moody algebras endowed with a loop structure is
one of the most outstanding properties shown by integrable nonlinear
field equations (NFE's). For integrable NFE's in 1+1 dimensions,
algebras of this type emerge in a natural way within the "direct"
Estabrook-Wahlquist
prolongation method \cite{estabrook-wahlquist}. Such algebras, which
can be regarded as
infinite-dimensional realizations of incomplete prolongation Lie
algebras
(in the sense that not all of the commutators are known), are derived
{\it via} the introduction of an arbitrary number of prolongation
forms
having new dependent variables (called pseudopotentials), by assuming
the algebraic equivalence between the generators of the prolonged
ideals and their exterior differentials. Conversely,
in 2+1 dimensions loop
algebras were found applying the symmetry reduction procedure (a
group-theoretical analysis \cite{olver}), rather than the prolongation
technique, to some
integrable NFE's of physical interest, such as the Kadomtsev-
Petviashvili \cite{david}
(KP) and the Davey-Stewartson (DS) equation \cite{champagne}, the 3WRI
system \cite{martina}, and
the Ishimori model \cite{profilo}.
Indeed, the prolongation theory in 2+1 independent variables (and,
generally, in higher dimensions) is a difficult task and so far only a
very few papers have been written on this subject \cite{morris}.

The authors of Refs. [3-6] derived essentially the incomplete
prolongation
Lie algebra associated with certain integrable NFE's in 2+1 dimensions.
In some  cases \cite{dodd} a quotient algebra is used to find the
linear eigenvalue
problem related to the equations under investigation. Anyway, the
techniques devised are not algorithmic and do not seem to be easily
implementable. Many problems remain open, such as that of building up
 a Kac Moody realization with a loop structure of the incomplete
prolongation
Lie algebras. At this stage we observe that the symmetry algebra
(namely, the algebra arising from the symmetry reduction procedure)
possessed by an
integrable NFE in 2+1 dimensions (KP, DS, ...), is of the Kac-Moody
type with a loop structure. Therefore, it seems reasonable to ask
whether a link
exists between the symmetry approach, which is based on the Lie group
theory, and the prolongation method, which is formulated in the
language of
differential forms \cite{estabrook-wahlquist}.
This is an important methodological aspect which
deserves a wide investigation. Here we remark only that any approach
to this problem should not ignore the contribution by Harrison and
Estabrook \cite{harrison}, where the fundamental concepts of Cartan's
theory of systems of partial
differential equations are exploited for obtaining  the generators of
their invariance groups (isogroups).

Another important aspect inherent to the integrability properties of
NFE's, is the possibility of applying the so-called "inverse"
prolongation method. This consists in starting from a given incomplete
Lie algebra to
generate the class of NFE's whose prolongation structure it is.
Interesting results in this direction have been obtained in Refs.
\cite{estabrook,hoenselaers,leo,alfinito}.

On the basis of our present knowledge, an  algebraic theory of
integrability implies at least the execution of two research programs,
i.e. i) the
program of founding on a sound ground the prolongation theory in more
than 1+1 dimensions and its possible connection with the symmetry
approach;
 ii) the program of constructing and using ({\it via} the inverse
prolongation method) Kac-Moody realizations of the loop type of the
incomplete prolongation Lie algebras. It is convenient to subordinate
the program i) to the second one, ii). This choice is motivated by the
fact
that, quite recently, within the inverse prolongation scheme a new
procedure has been outlined \cite{alfinito} for generating a class of
NFE's starting
from a loop algebra realization of a certain incomplete prolongation
Lie algebra. This procedure is based on an {\sl ansatz } which may be
generalized, in principle, to the case of more space variables.

{}From the above considerations it turns out that a close correspondence
can be established between loop algebras and integrability. However,
in order
to achieve a deeper understanding of this correspondence, we have to
deal with new case studies, possibly resorting to different approaches.
Following this idea, in this Letter we study a couple of nonlinear
Schr{\"o}dinger (NLS) equations in the direct and inverse
prolongation framework. This system
describes the propagation of waves in nonlinear birefringent optical
fibers.

The direct prolongation method provides an incomplete Lie algebra
which is used to find the linear eigenvalue problem associated with
the system of NLS
equations. Furthermore, we devise an inverse prolongation technique
which is based on a loop algebra realization of the incomplete
prolongation Lie
algebra, for obtaining a whole family of NLS pair of equations
containing
the original ones.

\section{The incomplete prolongation Lie algebra}

Let us consider the pair of NLS equations \cite{logvin}:
$$i \,u_{x}\,+\,u_{tt}\,+\,\kappa v\,+\,(\alpha\,|u|^{2}\,+\,\beta\,
|v|^{2})\,u\,=\,0, \eqno(2.1a)$$
$$ i\,v_{x}\,+\,v_{tt}\,+\,\kappa u\,+\,(\alpha\,|v|^{2}\,+\,\beta\,
|u|^{2})\,v\,=\,0, \eqno(2.1b)$$
where subscripts stand for partial derivatives,
 $u=u(x,t)$ and $v=v(x,t)$ are the circularly polarized components of
the optical field, $x$ and $t$ denote the (normalized) longitudinal
coordinate of the fiber and the time variable, respectively, $\kappa$
is
the birefringent parameter and the coefficients $(\alpha $ and $\beta$
are
responsible for the nonlinear properties of the fiber .

First, let us assume that $\kappa,\;\alpha$ and $\beta$ are
real parameters,
 with $\alpha=\beta$. Then, Eq.(2.1) can be written in the form
$$i\,\vec{u}_{x}\,+\,\vec{u}_{tt}\,+\,\kappa\,\sigma \,
\vec{u}\,+\,\alpha\,|\vec{u}|^{2}\,\vec{u}\,=
\,0, \eqno(2.2a)$$
$$-i\,\vec{u}^{*}_{x}\,+\,\vec{u}^{*}_{tt}\,+\,\kappa\,\sigma \,
\vec{u}^{*}\,+\,\alpha\,|\vec{u}|^{2}\,\vec{u}^{*}\,=
\,0, \eqno(2.2b)$$
where
$$ \sigma\,=\,\left(
\begin{array}{cc}
0 & 1 \\
1 & 0 \\
 \end{array}\right), \quad
\vec{u}\,=\,\left(
\begin{array}{c}
u  \\
v  \\
\end{array}\right), \quad
\vec{u}^{*}\,=\,\left(
\begin{array}{c}
u^{*}  \\
v^{*} \\
\end{array}\right), \quad |\vec{u}|^{2}\,=\,\vec{u}\cdot
\vec{u}^{*}$$
(the dot means the scalar product).

The EW prolongation for Eqs. (2.2) can be formulated by introducing the
differential ideal defined by the set of (vector) 2-forms:
$$\vec{\theta}^{1}\,=\,\de \vec{u} \wedge \de x\,-\, \vec{u}_{t}\,\de t
\wedge \de x, \eqno(2.3a)$$
$$\vec{\theta}^{2}\,=\,d \vec{u}^{*} \wedge d x\,-\,
\vec{u}^{*}_{t}\,d t\wedge \de x ,\eqno(2.3b)$$
$$\vec{\theta}^{3}\,=\,-i\,\de \vec{u} \wedge \de t\,+\,\de
\vec{u}_{t}\wedge \de x\,+\,\kappa\,\sigma\, \vec{u} \de t\wedge\de x\,+\,
\alpha\,| \vec{u}|^{2}\, \vec{u}\,\de t\wedge \de x ,\eqno(2.3c)$$
$$\vec{\theta}^{4}\,=\,i\,\de \vec{u}^{*} \wedge \de t\,+\,\de
\vec{u}^{*}_{t}\wedge \de x\,+\,\kappa\,\sigma\, \vec{u}^{*}\de t
\wedge\de
x\,+\,\alpha\,| \vec{u}|^{2}\, \vec{u}\,\de t\wedge \de x.
\eqno(2.3d) $$
One can verify that the ideal (2.3) is closed.

Now we look for the prolongation 1-forms :
$$\omega^{k}\,=\, \de y^{k}\,+\, F^{k}(\vec{u},\vec{u}^{*},
\vec{u}_{t}
,\vec{u}^{*}_{t};y)\de x\,+\,G^{k}(\vec{u},\vec{u}^{*};y)\de t,
\eqno(2.4) $$
where $y=\{y^{m}\}$ is the pseudopotential, ($k,\,m\,=\,1,2,\ldots,
{\rm N}$ (N arbitrary)), and $F^{k}$ and $G^{k}$ are functions to be
determined. (In the following we shall drop the index $k$, for
simplicity).
The requirement that the forms $\vec{\theta}^{j}\;\;(j=1,\ldots,4)$
and
$\omega^{k}$ comprise a differential ideal
${\cal I}(\theta^{j},\omega^{k}$), i.e. d$\omega^{k} \;\;\in \;
 {\cal I}(\theta^{j},\omega^{k}$), implies the set of constraints
$$ i F_{\vec{u}_{t}}\,+\,G_{\vec{u}}\,=0, \eqno(2.5a)$$
$$ i F_{\vec{u}^{*}_{t}}\,-\,G_{\vec{u}^{*}}\,=0, \eqno(2.5b)$$
$$ F_{\vec{u}}\,\cdot\,\vec{u}_{t}\,+\, F_{\vec{u}^{*}}\,\cdot\,
\vec{u}^{*}_{t}\,
-\,\kappa\, F_{\vec{u}_{t}}\,\cdot\,\sigma\,\vec{u}\,-\,\alpha\,
|\vec{u}|^{2}
\, F_{\vec{u}_{t}}\,\cdot\,\vec{u}\,-\,
\kappa\,F_{\vec{u}^{*}_{t}}\,\cdot\,\sigma\,\vec{u}^{*}\,$$
$$-\,\alpha\,|\vec{u}|^{2}\,F_{\vec{u}^{*}_{t}}\,\cdot\,\vec{u}^{*}\,
+\,\left[F,G\right]\,=\,0,\eqno(2.5c)$$
where
$$ F_{\vec{u}}\,=\,{\rm grad}_{\vec{u}}\,F\,\equiv\,(F_{u},F_{v}),$$
the symbol $\left[F,G\right]$ stands for $\left[F,G\right]^{k}\,=\,
F_{j}\,\frac{\partial G^{k}}{\partial y_{j}}\;-\;G_{j}\,
\frac{\partial F^{k}}{\partial y_{j}},$ and $F_{u}\,=\,
\frac{\partial F}{\partial u}$, and so on.

{}From (2.5) we obtain
$$\begin{array}{ll}
 -i F\,=& \,\,\vec{a}\cdot\vec{u}_{t}\,+\,\vec{u}^{*}\cdot
M \vec{u}_{t}
\,-\,\vec{b}\,\cdot\,\vec{u}^{*}_{t}\,-\,\vec{u}^{*}_{t}\cdot
M \vec{u}\, \\
&-\, [\vec{a}\cdot \vec{u},\,\vec{b}\cdot \vec{u}^{*}]\,-\,
[\vec{a}\cdot\vec{u},\,c]\,+\,[\vec{b}\cdot\vec{u}^{*},\,c]\,
-i d,\\ \\
\end{array}\eqno(2.6a)$$
$$G\,=\,{\vec a}\cdot{\vec u}\,+\,{\vec b}\cdot{\vec u}^{*}\,+\,
{\vec u}^{*}\cdot M{\vec u}\,+\, c, \eqno(2.6b)$$
and the commutation (Lie brackets) relations
$$[a_{j},a_{k}]=\,0, \quad [b_{j},b_{k}]=\,0,\quad [a_{j},M_{kl}]=
\,0,$$
$$[b_{j},M_{kl}]=\,0,\quad [M_{jk},M_{lm}]=\,0,\quad [c,M_{jk}]=
\,0,$$
$$[c,d]=\,0, \quad [[a_{j},c],a_{k}]=\,0,\quad [[b_{j},c],b_{k}]=
\,0,$$
$$i \kappa \sigma {\vec a}\,+\,i [[{\vec a},c],c]\,-\,[d,{\vec a}]\,=
\,0,$$
$$i \kappa \sigma {\vec b}\,+\,i [[{\vec b},c],c]\,+\,[d,{\vec b}]\,=
\,0,\eqno(2.7)$$
$$[[b_{k},a_{l}],a_{j}]=\,\frac{1}{2}\alpha(\delta_{kl}a_{j}\,+\,\delta_{kj}
a_{l}),$$
$$[[b_{k},a_{l}],b_{j}]=\,-\frac{1}{2}\alpha(\delta_{kl}b_{j}\,+\,\delta_{lj}
b_{k}),
$$
$$-i \kappa\sigma (M\sigma-\sigma M)_{jk}\,+\,2i [[b_{j},a_{k}],c]\,
+\,[d,M_{jk}]\,=\,0.$$
In (2.6) and (2.7), $M$ is a $2\times 2$ matrix whose elements
$M_{jk}$, together with the functions $c,\,d$ and the vectors
${\vec a}\equiv
(a_{1},a_{2}),\;{\vec b}\equiv (b_{1},b_{2})$, depend arbitrarily on
the pseudopotential only. The indices $j,k,l,m$ take the values 1,2.
Furthermore, one has
$$ M^{\dag}\,=\,M,\quad {\vec u}^{*}\cdot M{\vec u}\,=\,
{\vec u}\cdot M^{\rm T}{\vec u}^{*} \eqno(2.8)$$
where T means transposition.

The prolongation algebra (2.7) turns out to be incomplete. However, as
we shall see in Sections 3 and 4, some important properties of Eqs.
(2.2)
 arise from an infinite-dimensional realization of the Kac-Moody type
with a loop structure.

\section{Quotient algebra and spectral problem}
For practical purposes, say to find the spectral problem associated
with Eqs.(2.2) and, possibly, B{\"a}cklund transformations, the
algebra (2.7) can be closed to provide finite-dimensional quotient
algebra. This can be done assuming, for example, that
$$ M\sigma\,=\,\sigma M,\qquad [d,M_{jk}]\,=\,0, \eqno(3.1)$$
the quantities $a_{1},\,a_{2},\,b_{1},\,b_{2}$, $[a_{1},b_{1}]\equiv
t_{11},$
$[a_{1},b_{2}]\equiv t_{12},$ $[a_{2},b_{1}]\equiv t_{21},$ $[a_{2},
b_{2}]\equiv t_{22}$
are independent, and $c$ and $d$ are a linear combination of the
preceding ones. In doing so, we get (see (2.8) and (3.1))
$$M\,=\,\left(\begin{array}{cc} q & r\\ r & q\\
\end{array}\right),$$
$q$ and $r$ being two real functions of the pseudopotential;
$$c\,=\,\lambda(t_{11}\,+\,t_{22}),\eqno(3.3)$$
$$d\,=\, -3i \lambda c\,-\,i \kappa\sum_{j,k=1,2}(1-\delta_{jk})
t_{jk}, \eqno(3.4)$$
and
$$[a_{j},a_{k}]\,=\,0,\quad [b_{j},b_{k}]\,=\,0,\quad
[a_{j},b_{k}]\,=\,t_{jk},\eqno(3.5)$$
$$[a_{k},t_{lm}]\,=\,a_{k}\delta_{lm}\,+\,a_{l}\delta_{km},
\quad [b_{m},t_{lk}]\,=\,-(b_{k}\delta_{lm}\,+\,b_{m}\delta_{lk}), $$
$$[t_{jk},t_{lm}]\,=\,t_{lk}\delta_{jm}\,-\,t_{jm}\delta_{kl}
\quad (j,k,l,m \,=\,1,2).$$
In (3.3), (3.4) and (3.5) we have performed the changes ${\vec a}
\rightarrow\frac{1}{2} \alpha {\vec a},\;\;\lambda\rightarrow
\frac{2}{\alpha}\,
\lambda$, where $\lambda$ is a free parameter. The commutation
relations
(3.5) define an sl(3,$\cal C$) quotient algebra of the incomplete
prolongation algebra (2.7). We notice that exploiting (3.5), (2.4) and
(2.6), the spectral problem related to Eqs.(2.2) is furnished.

Another interesting realization of (2.7) is given by an infinite-
dimensional Lie algebra of the Kac-Moody type.
This realization can be built up by putting $M\,=\,0$ in (2.7) and
resorting to the correspondence
$$ a_{m}\rightarrow\,\T_{0m}^{(n)},\quad
b_{m}\rightarrow\,\T_{m0}^{(-n)},$$
$$ t_{lm}\rightarrow\,i \T_{lm}^{[n(\delta_{l0}\,-\,\delta_{m0})]},
\quad
 c\rightarrow\,\lambda\left(\T_{11}^{(l)}\,+\,\T_{22}^{(l)}
\right),\eqno(3.6)$$
$$
d\rightarrow\,-3i\,\lambda^{2}\left(\T_{11}^{(2l)}\,
+\,\T_{22}^{(2l)}\right)\,-i\,\kappa\left(\T_{12}^{(0)}\,+\,
\T_{21}^{(0)}\right),$$
where $n,l \in \cal Z$.

The vector fields $\T_{lm}^{(n)}$ satisfy the commutation relations
$$\left[\T_{lm}^{(n)},\T_{l^{'} m^{'}}^{(n^{'})}
\right]\,=\,i \epsilon_{lml^{'} m^{'} kj}\,\T_{kj}^{(n+n^{'})}
\eqno(3.7)$$
where the structure constants are expressed by
$$\begin{array}{ll}
\epsilon_{lml^{'} m^{'} kj}\,=&\,\delta_{lm}\delta_{l^{'} 0}
\delta_{k0}\delta_{m^{'} j}\,-\,\delta_{lm}\delta_{m^{'} 0},
\delta_{j0}\delta_{l^{'} k}\\
& +\,\delta_{l^{'} m^{'}}\delta_{m0}\delta_{nl}\delta_{j0}\,-\,
\delta_{l^{'} m^{'}}\delta_{m0}\delta_{kl}\delta_{j0}\\
& +\, \delta_{m^{'} l}\delta_{kl^{'}}\delta_{jm}\,-\,
\delta_{m l{'}}\delta_{kl}\delta_{jm^{'}} \\
\end{array}
\eqno(3.8)$$
$(m,l,m^{'}, l^{'}, j,k\,=\, 0,1,2). $
We observe that $\T^{(n)}_{00}=0$, $\forall n \in {\cal Z}$.

If, in particular, $l,m,l^{'}, m^{'}\,\ne\,0$, Eq.(3.7) becomes
$$\left[ \T_{lm}^{(n)},\T_{l^{'} m^{'}}^{(n^{'})}\right]\,=\,
i \delta_{m^{'} l}\T_{l^{'} m}^{(n+n^{'})}\,-\,i
 \delta_{ml^{'}}\T_{lm^{'}}^{(n+n^{'})}.\eqno(3.9)$$
It is noteworthy that (3.7) admits a representation in terms
of the prolongation variables, namely
$$\T_{lm}^{(r)}\,=\,- i \sum^{+\infty}_{k=-\infty}\,
y^{(r+k)}_{l}\,\frac{\partial}{\partial y_{m}^{(k)}}\,+\,
i \delta_{lm}\,\sum^{+\infty}_{k=-\infty}
y^{(r+k)}_{0}\,\frac{\partial}{\partial y_{0}^{(k)}},\eqno(3.10)$$
$r \in {\cal Z}$,
where the pseudopotential $y$ is expressed through the independent
infinite-dimensional vectors $y_{0},\;y_{1}$ and $y_{2}$
($y_{0}^{(k)},\;
y_{1}^{(k)}$ and $y_{2}^{(k)}$ denote the $k$-component of $y_{0},\;
y_{1}$ and $y_{2}$, respectively).

Now we are ready to derive the spectral problem for Eqs.(2.2) using
the Kac-Moody loop algebra (3.7). To this aim, from (2.4) and (2.6)
(with $M=0$) we obtain
$${\vec \Psi}_{x}(\lambda)\,=\, L_{1}\,{\vec \Psi}(\lambda),
\eqno(3.11a) $$
$${\vec \Psi}_{t}(\lambda)\,=\, L_{2}\,{\vec \Psi}(\lambda),
\eqno(3.11b) $$
where ${\vec \Psi}$ is a 3-component vector such that
$$ {\vec \Psi}(\lambda)\,\equiv\,{\vec \Psi}(x,t;\lambda)\,=\,
\sum^{+\infty}_{k=-\infty}\,\lambda^{k}{\vec y}^{(k)}, \eqno(3.12)$$
${\vec y}^{(k)}\,=\,( y_{0}^{(k)},\,y_{1}^{(k)},\,
y_{2}^{(k)})^{\rm T}$,
and $L_{1}$ and $L_{2}$ are the $3\times 3$ traceless matrices
$$L_{1}\,=\,\left(
\begin{array}{lll}
-i(\frac{1}{2}\alpha|u|^{2}\,+\,\frac{1}{2}\alpha|v|^{2}\,+\,6\lambda^{2})\quad
 &
\quad
u^{*}_{t}\,+\,3\lambda u^{*}\quad &\quad  v^{*}_{t}\,+\,3\lambda
v^{*}\\
\frac{1}{2}\alpha(-u_{t}\,+\,3\lambda u)\quad  &\quad i\,(\frac{1}{2}\alpha
|u|^{2}\,
+\,3 \lambda^{2})\quad &\quad i\,(\frac{1}{2}\alpha v^{*}u\,+\,\kappa)\\
\frac{1}{2}\alpha (-v_{t}\,+\,3\lambda v)\quad &\quad  i\,(\frac{1}{2}\alpha v
u^{*}\,
+\,\kappa)\quad &\quad
i\,(\frac{1}{2}\alpha |v|^{2}\,+\,3\lambda^{2})\\
\end{array}\right), \eqno(3.13)$$
$$L_{2}\,=\,\left(
\begin{array}{lll}
2\lambda \quad &\quad  i u^{*}\quad & \quad i v^{*}\\
 i \frac{1}{2} \alpha u \quad &\quad -\lambda \quad &\quad  0 \\
 i \frac{1}{2} \alpha v \quad &\quad 0 \quad & \quad-\lambda \\
\end{array}\right). \eqno(3.14)$$

The compatibility condition for Eqs.(3.11), i.e. $L_{1t}\,-\,L_{2x}\,
+\,[L_{1},L_{2}]\,=\,0$, reproduces just the equations (2.2).

\section{Incomplete Lie algebras vs. nonlinear field equations}

The prolongation of a nonlinear field equation can be interpreted as a
Cartan-Ehresmann connection. In this way an incomplete Lie algebra of
vector fields can be related to a differential ideal
\cite{estabrook}. Here we start
from the incomplete Lie algebra (2.7) to yield the differential ideal
associated with Eqs.(2.2). This can be carried out specifying the form
of the connection. In doing so, let us assume that the connection
$$ \omega^{k}\,=\,\de y^{k}\,+\,A^{k}_{j}\eta^{j}\eqno(4.1)$$
exists such that
$$\de\omega^{k}\,=\,A^{k}_{j}\de\eta^{j}\,-\,\frac{1}{2}\left[A_{i},A_{j}
\right]^{k}\,\eta^{i}\wedge\,\eta^{j}\,=\,0\qquad ({\rm mod}\;
\omega^{k}),\eqno(4.2)$$
where $\eta^{j}$ are 1-forms, $A_{j}$ (j=1,2,$\ldots$, 18) are defined
by
$$A_{1}\,=\,a_{1},\; A_{2}\,=\,a_{2},\; A_{3}\,=\,b_{1},\;
A_{4}\,=\,b_{2},\; A_{5}\,=\,c\; $$
$$A_{6}\,=\,d,\; A_{7}\,=\,t_{11},\; A_{8}\,=\,t_{12},\;
A_{9}\,=\,[a_{1},c],\; A_{10}\,=\,t_{21},\;$$
$$ A_{11}\,=\,t_{22},\;A_{12}\,=\,[a_{2},c],\,
A_{13}\,=\,[b_{1},c],\; A_{14}\,=\,[b_{2},c],\;
A_{15}\,=\,M_{12}$$
$$ A_{16}\,=\,M_{21},\; A_{17}\,=\,M_{11},\; A_{18}\,=\,M_{22},\;$$
involved in the incomplete Lie algebra (2.7), and mod $\omega^{k}$
means that all the exterior products between $\omega^{k}$ and 1-forms
of the Grassmann algebra have not been considered.

By expliciting (4.2) and taking into account (2.7), we get the set of
exterior differential equations
$$\de\eta^{1}\,-\,\kappa\eta^{5}\wedge\eta^{12}\,-\,\alpha\eta^{1}\wedge
\eta^{7}\,-\,\frac{1}{2}\alpha\left(\eta^{1}\wedge\eta^{11}\,+\,\eta^{2}\wedge
\eta^{8}\right)\,=\,0, \eqno(4.3a)$$
$$\de\eta^{2}\,-\,\kappa\eta^{5}\wedge\eta^{9}\,-\,\alpha\eta^{2}\wedge
\eta^{11}\,-\,\frac{1}{2}\alpha\left(\eta^{1}\wedge\eta^{10}\,+\,\eta^{2}\wedge
\eta^{7}\right)\,=\,0, \eqno(4.3b)$$
$$\de\eta^{3}\,-\,\kappa\eta^{5}\wedge\eta^{14}\,+\,\alpha\eta^{3}\wedge
\eta^{7}
\,+\,\frac{1}{2}\alpha\left(\eta^{3}\wedge\eta^{11}\,+\,\eta^{4}\wedge\eta^{10}
\right)\,=\,0, \eqno(4.3c)$$
$$\de\eta^{4}\,-\,\kappa\eta^{5}\wedge\eta^{13}\,+\,\alpha\eta^{4}\wedge
\eta^{11}\,+\,\frac{1}{2}\alpha\left(\eta^{3}\wedge\eta^{8}\,+\,\eta^{4}\wedge
\eta^{7}\right)\,=\,0, \eqno(4.3d)$$
and another set of constraints, which we omit for brevity, whose
solutions
can be written as
$$\eta^{j}\,=\,-\,\alpha^{j}_{t}\de x\,+\,i \alpha^{j} \de t\quad
(j=1,2),$$
$$\eta^{k}\,=\,-\,\alpha^{k}_{t}\de x\,-\,i \alpha^{k} \de t
\quad (k=3,4),$$
$$\eta^{5}\,=\,\de t,\,\quad \eta^{6}\,=\,\de x,$$
$$\eta^{7}\,=\,-\,i \alpha^{1}\alpha^{3} \de x,\quad \eta^{8}\,=\,-\,
i \alpha^{1}\alpha^{4} \de x,\quad \eta^{9}\,=\, \alpha^{1}
\de x,\eqno(4.4)$$
$$\eta^{10}\,=\,-\,i \alpha^{2}\alpha^{3} \de x,\eta^{11}\,=\,-\,i
\alpha^{2}\alpha^{4} \de x,\,\eta^{12}\,=\, \alpha^{2} \de x,$$
$$\eta^{13}\,=\, \alpha^{3} \de x,\quad \eta^{14}\,=\,\alpha^{4} \de x,$$
where $\alpha^{l}\quad (l\,=\,1,\ldots,4)$ are arbitrary 0-forms. At this
stage, by choosing $\alpha^{1}\,=\,u,\quad\alpha^{2}\,=\,v,\quad
\alpha^{3}\,=\,u^{*},\quad\alpha^{4}\,=\,v^{*},$ by virtue of (4.4), Eqs.
(4.3) give exactly the original system (2.2).

In the following we solve the inverse prolongation problem for
Eqs.(2.2) by adopting a procedure based on the Kac-Moody
realization
(3.10). Precisely, let us look for the class of NFE's whose
prolongation structure is assumed to be given by the 1-forms
$$ \omega \,=\, \de y\,+\, {\cal F}({\vec u},{\vec u}^{*},{\vec u}_{t},
{\vec u}^{*}_{t};y)\de x
\,+\,{\cal G}({\vec u},{\vec u}^{*};y)\de t, \eqno(4.5)$$
where ${\cal F}$ and  ${\cal G}$ are defined by
$$ {\cal F}\,=\,-i \sum^{2}_{k=1}\,\left[
\alpha_{k} \T^{(n)}_{0k}\,+\,\beta_{k} \T^{(-n)}_{k0}\,+\,\gamma_{k}
\T^{(n+l)}_{0k}\,+\,\phi_{k} \T^{(-n+l)}_{k0}\,
+\,\psi_{k} \T^{(0)}_{1k}\,\right . $$
$$+\left .\,\chi_{k} \T^{(0)}_{2k}\,+\,\rho \T^{(2l)}_{kk}\right],
\eqno(4.6a)$$
$$ {\cal G}\,=\,-i \sum^{2}_{k=1}\,\left[
p_{k} u_{k} \T^{(n)}_{0k}\,+\,q_{k} u^{*}_{k} \T^{(-n)}_{k0}\,
+\,\sigma \T^{(l)}_{kk}\right], \eqno(4.6b)$$
$\alpha_{k},\,\beta_{k},\,\gamma_{k},\,\phi_{k},\,\psi_{k},\,\chi_{k}, $
are functions of $u=u_{1},\,v=u_{2},\, u^{*}=u^{*}_{1},
\,v^{*}=u^{*}_{2}$ to be determined in such a way that
the operators $\T (\cdot)$ satisfy the Kac-Moody algebra (3.7) and
$ p_{k},\,q_{k},\,\rho,$ and $\sigma$ are constants.

At this point, by setting $\omega \,=\,0$ into Eqs. (4.5) from the
compatibility condition ${\vec y}^{(i)}_{xt}\,=\,{\vec y}^{(i)}_{tx}$
we obtain the constraints
$$\beta_{1t}\,-\, \left(\psi_{1}+\chi_{2}\right)q_{1}u^{*}_{1}\,=\,
\sum^{2}_{j=1}\,q_{j}\psi_{j}u^{*}_{j}\,+\,q_{1}u^{*}_{1x},\eqno(4.7)$$
$$\beta_{2t}\,-\, \left(\psi_{1}+\chi_{2}\right)q_{2}u^{*}_{2}\,=\,
\sum^{2}_{j=1}\,q_{j}\chi_{j}u^{*}_{j}\,+\,q_{2}u^{*}_{2x},\eqno(4.8)
$$

$$\alpha_{k,t}\,+\,p_{k}u_{k}\left(\psi_{1}+\chi_{2}\right)\,
+\,p_{1}u_{1}
\psi_{k}\,+\,p_{2}u_{2}\chi_{k}\,=\,p_{k}u_{k,x}
\eqno(4.9)$$
and
$$\alpha_{k}\,=\,\frac{\rho}{3\sigma^{2}}\,p_{k}u_{kt},\quad
\beta_{k}\,=\,-\,\frac{\rho}{3\sigma^{2}}\,q_{k}u^{*}_{kt},$$
$$\gamma_{k}\,=\,\frac{\rho}{\sigma}\,p_{k}u_{k},\quad
\phi_{k}\,=\,\frac{\rho}{\sigma}\,q_{k}u^{*}_{k},$$
$$\psi_{k}\,=\,-\,\frac{\rho}{3\sigma^{2}}\,q_{1}p_{k}u^{*}_{1}u_{k}
\,+\,\eta_{k},$$
$$\chi_{k}\,=\,-\,\frac{\rho}{3\sigma^{2}}\,q_{2}p_{k}u^{*}_{2}u_{k}
\,+\,\mu_{k},\eqno(4.10)$$
where $\eta_{k}$ and  $\mu_{k}$ are arbitrary constants and $k=1,2$.
Since $\frac{3\sigma^{2}}{\rho}$ is an imaginary number,
$p_{k}q_{k}$ is a real constant, and $\eta_{2}=-\mu^{*}_{1}$,
$\mu_{2}=-\mu^{*}_{2}$, $\eta_{1}=-\eta^{*}_{1}$, without lost
of generality we can put $\rho\,=\,3i \sigma^{2}$, $p_{1}\,=\,p_{2}
\,=i$, $q_{1}\,=\,q_{2}\,=\frac{i}{2}\,\epsilon $, where
$\epsilon$ is
a real quantity. Then, with the help of (4.10), Eqs.(4.7)-(4.9) yield
$$i\,{\vec u}_{x}\,+\,{\vec u}_{tt}\,+\,A \,
{\vec u}\,+\,\epsilon\,|{\vec u}|^{2}\,{\vec u}\,=\,0, \eqno(4.11a)$$
$$-i\,{\vec u}^{*}_{x}\,+\,{\vec u}^{*}_{tt}\,+\,A^{\rm T} \,
{\vec u}^{*}\,+\,\epsilon\,|{\vec u}|^{2}\,{\vec u}^{*}\,=\,0,
\eqno(4.11b)$$
where
$$A\,=\,\left(
\begin{array}{ll}
\Delta_{1}\;\;& \;\;\kappa \\
 \kappa^{*}\;\;&\;\;\Delta_{2}
\end{array}\right),\eqno(4.12)$$
$\Delta_{1}\,=\,-\,i \left(2\eta_{1}\,+\,\mu_{2}\right)$
$\in {\cal R}$,
$\Delta_{2}\,=\,-\,i \left(\eta_{1}\,+\,2\mu_{2}\right)$
$\in {\cal R}$,
$\kappa\,=\,-\,i\,\mu_{1}$ $\in {\cal C}$, and $A^{\rm T}\,=\,A^{*}$.

We remark that the nonlinear field equations obtained by our method,
i.e.
Eqs.(4.11), are more general than Eqs.(2.2). These can be found for
$\Delta_{1}\,=\,\Delta_{2}\,=0$ and $\kappa\,=\,\kappa^{*}$.
Furthermore, we
point out that our inverse prolongation technique, based on the
{\sl ansatz} (4.5) which exploits the Kac-Moody realization (3.10),
is a powerful
tool for generating new integrable nonlinear field equations of
physical significance. In fact, in the present case, by choosing
$\Delta_{1}\,=\,-\,\Delta_{2}\,=\Delta \ne 0$ and $\kappa\,=\,
\kappa^{*}$,
 Eqs.(4.11) become the coupled equations related to twisted
birefringent optical fibers \cite{trillo}.

\section{Conclusions}
The results achieved in this Letter show that the Estabrook-Whalquist
prolongation method presents several advantages in the study of
nonlinear field equations. The method works out both in the "direct"
and in the "inverse" direction. Our investigation confirms the
existence of a deep connection between prolongation Lie algebras and
the integrability
property of nonlinear field equations. In particular, concerning the
nonlinear Schr{\"o}dinger equations (2.2), we point out that the
related incomplete Lie algebra (2.7) admits an infinite-dimensional
realization of the Kac-Moody type which yields the linear spectral
problem associated with
the system under consideration. On the other hand, it is noteworthy
that the same Kac-Moody algebra is involved in the inverse
prolongation procedure. This, which is based on the {\sl ansatz}
(4.5), produces the new class of linearizable nonlinear field
equations (4.11), containing the starting ones, i.e. Eqs. (2.2),
as a particular case. In this context, we
have that for $\Delta_{1}=-\Delta_{2}\equiv \Delta \ne 0$ and
$\kappa=\kappa^{*}$,
Eqs.(4.11) coincide with the system describing twisted birefringent
optical fibers \cite{trillo}. This result is interesting mostly in
the sense that
so far the integrability property of such equations was not known.

Let us conclude with a few comments. First, the prolongation method
possesses
an intrinsic predictive character. In other words, in the specific
application carried out here, the prolongation algebra related to
Eqs.(2.1) allows a nontrivial closure if and only if $\alpha \equiv
\beta
\in {\cal R}$. Second, although some examples of prolongation
calculations exist in the direct framework in more than $1+1$
dimensions \cite{morris}, it seems that, on the contrary,
the inverse method has not been still explored.
This is an open problem, together with the attempt of building up an
algebraic foundation of integrability of nonlinear field equations.
The prolongation strategy might accomplish this plan.

\end{document}